# Consequence of doping in spatiotemporal rock-paper-scissors games

Wang Zhijian

Interdiscriptional Center of Social Science, Zhejiang University, 310027, China

email: wangzj@zju.edu.cn

**Abstract**

What determines species diversity is dramatic concern in science. Here we report the effect of doping on diversity in spatiotemporal rock-paper-scissors (RPS) games, which can be observed directly in ecological, biological and social systems in nature. Doping means that there exists some buffer patches which do not involve the main procession of the conflicts but occupied the game space. Quantitative lattices simulation finds that (1) decrease of extinction possibility is exponential dependent on the increase of doping rate, (2) the possibility of the conflict is independent of doping rate at well mix evolution beginning, and is buffered by doping in long time coexistence procession. Practical meaning of doping are discussed. To demonstrate the importance of doping, we present one practical example for microbial laboratory efficient operation and one theoretical example for human-environment co-existence system better understanding. It suggests that, for diversity, doping can not be neglected in spatial RPS games study and in practical operation.

## Introduction

Doping is widely existence in nature. There always exists some special pattern in a system, which do not involve the main procession but occupied the game spatial space and effects the systems performance. For example, doping leads material systems performance fancying, and without doping, there is no high temperature superconductivity, nor semiconductor. In social Dilemma, the way to cooperation is narrow [1]. However, doping, due to the loner who seems inactive, can lead co-existence of diversity of defect, cooperation and Tat-for-Tit instead of collapsing to dull defection world.[2] In ecology, several hundreds years ago Darwin found that spatial bar due to geographical splitting hold diversity of biology. In Ethnic/Cultural systems, doping, due to well mix and no ultra population demographical patches or geographic splitting, help to peace keeping[3]. So, intuitively, doping is a ubiquitous feature of real natural and social systems. In this letter, we study doping in the spatial evolutionary game theory.

Evolutionary spatial game theory provides a framework to investigate a broad class of co-development of subpopulations theoretically. In this context, the rock–paper–scissors game has emerged as a paradigm to describe species diversity, and exhibiting such dynamics have been identified in numerous ecosystems [4,5,6,7,8,9,10] . For example, microbial laboratory spatial structure on time development and coexistence of species[11]. Theoretically, spatiotemporal RPS games, also named as cyclic Lotka-Volterra model,[12] is studied with simulations on the lattice, mean-field approximation, pair approximation and beyond[15] . Even though spatiotemporal pattern in nonequilibrium systems [13] and evolutionary games on graphs[14] and spatial aspects of inter specific competition[15] are well studied, e.g. observed in some spatial three-strategy evolutionary public good games and Prisoner's Dilemma games. Just recently, Reichenbach, Mauro Mobilia and Frey [16] found that diversity drastically depend on mobility. However the effect due to random spatial doping on RPS is ignored as dilutes [15], and is not quantitatively study.

We find that doping, as a ubiquitous feature of real

RPS like ecosystems, affects diversity drastically also. We present quantitatively the consequence of doping by stochastic distribution the loner/desert patches, or white patches in the game RPS space. In the following we first survey the picture that can be deduced from Monte Carlo lattices simulations of the spatiotemporal evolutionary RSP games. The results will be compared with the general mean-field approximation and doping ignore model. The quantitative result is that (1) possibility of extinction decrease is exponential dependent on the doping rate increase, (2) the possibility of the conflict is long time keeping among the RPS subpopulation and being doping rate independent. The exponential relations mean that doping is not ignorable. Scientific meaning of doping in environment protection, ethnic/cultural conflict buffer and epidemiology will be discussed. As identified examples, we show briefly the practically usages of our result in microbial laboratory experiment and the usage on theoretical understanding the co-existence of ecological diversity and human population. As a main result, we suggest that doping effect can not be neglected in practically understanding diversity.

## Model and Method

First, consider the game space. It is the square lattice with von Neumann neighborhood (including connections between nearest 4 sites neighbor). We assume the space is doped by white patches, represent the loner or desert or river or mountain or bar like patches which occupied the game space, in the rate $\lambda$,. The white patches do not involve the main procession of the competition of the game. When the white patch do not involve the game at all, it was named as diluted lattices[17],also, which can be used to study what happens if some of lattices are removed at random. However, practically, the patches affect the system, for example, when people settlement on a new area, it effects on the local ecologic system, even thought it do not prey any spices, it might be nutrient and destroy the balance within the subpopulation. We do not pick the doping lattices out. In this letter, we show the buffer affect also.

Second, consider individuals of RPS three subpopulations (referred to as A, B, and C), arranged on a spatial lattice, where they can only interact with nearest neighbors, selection (to prey) or reproduction (to breed). For the possible interactions, the version of the rock–paper–scissors game, namely a stochastic spatial variant of the model is introduced by May and Leonard in 1975[15], Hofbauer and Sigmund in1988, Tainaka 2001 [18], and Reichenbach, Mobilia and Frey 2006, 2007[17].

Schematic illustrations of the model's dynamics are provided in Fig. 1. The basic reactions comprise selection and reproduction processes, which occur at rates u and w, respectively. The white patches, doping, refer to as D, do not occupied the black patches and do not be preyed by or to prey any subpopulation either A, or B and or C.

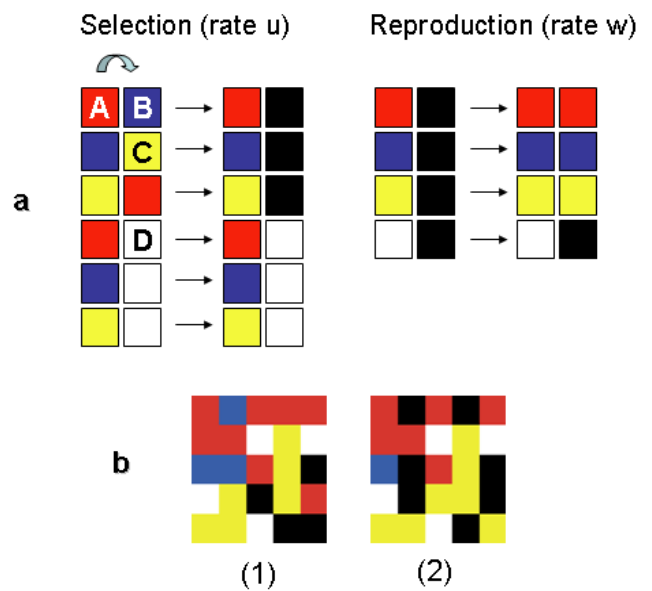

**Figure 1 | The rules of the stochastic model.** Individuals of three competing species A (red), B (blue), C (yellow) and D(white) occupy the sites of a lattice**. a,** They interact with their nearest neighbours through selection or reproduction, both of which reactions occur as same possibility processes at rates u and w, respectively. Selection reflects cyclic dominance: A can kill B, yielding an empty site (black). In the same way, B invades C, and C in turn outcompetes A. Reproduction of individuals is only allowed on empty neighbouring sites, to mimic a finite carrying capacity of the system. D do not interact with neighbours , and might act as buffer to enhance/reduce the rate ***u*** and/or ***w***. **b,** An example of the processes, taking place on a 3×3 square lattice. (2) is the next iteration of (1). Selection and reproduction procession present. The white patches, D, unchanging.

The main propose of this letter is to perform the consequence of doping; we take both the selection rate *u* and reproduction rate *w* to be 1 both. We performed extensive computer simulations of the stochastic system and typical snapshots of the evolution show in Fig. 2

## Result

**Diversity Phase Transformation & Doping Rate**

As doping is a ubiquitous feature of real systems, the quantitatively learning the consequence of doping is necessary. Here, we show the relation between the diversity-uniform phase transformations and doping rate, and find that the increase of diversity is exponential depend the increasing rate of doping.

We have considered the extinction probability $P_{ext}$ that, starting with randomly distributed individuals on a square lattice, the system has reached an absorbing state after a waiting time $t = N$. We compute $P_{ext}$ as a function of the doping rate (with $u$=1 and $w$=1). Notices that: $P_{ext}$ = (1 – Co-existence Possibility)

As Reichenbach, Mauro Mobilia and Frey in studying mobility effect on diversity, we consider also the typical waiting time $T$ until extinction occurs, and its dependence on the system size $N$. If $T(N) / N$, the stability of coexistence is marginal[see Ref.12, 16]. Conversely, longer (shorter) waiting times scaling with higher (lower) powers of N indicate stable (unstable) coexistence. These three scenarios can be distinguished by computing the probability $P_{ext}$ that two species have gone extinct after a waiting time $t / N$. (see attachment for more detail on the relation within extinct times and spatial size and doping rate $\lambda$)

In Fig. 3, the extinction probability $P_{ext}$ vs doping rate $\lambda$ show. It performs that the effect of doping on diversity-uniform, do not depend on the size of lattice sensitively. The increase of diversity exponentially depends on the increasing rate of doping.

It is not surprise for high doping region, where the white patch spilt the lattice due to the percolation phase transformation of buffer, so the diversity protected. However, with low doping rate, e.g., 10%, the possibility of extinct is quickly lowed, e.g. 40%. The empirical relation within the $P_{ext}$ and doping, from our numerical experiments, is as:

$$P_{ext} = \alpha - \beta \ln(1+\lambda) \quad \text{(Eq. 1)}$$

in which the parameter $\alpha = 0.1$ and $\beta = 0.45$ with linear fitting, when $\lambda < 0.25$. It is normal for diversity can always keeping, because of high $\lambda$ lead to white patch percolation phase transformation.

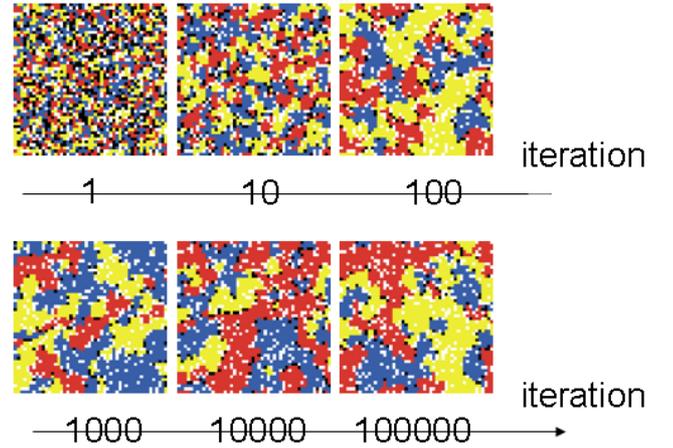

**Fig 2, Pattern evolution.** We show snapshots obtained from lattice simulations in 15% doping, a 50×50 two-dimensional square lattice with periodic boundary conditions. Typical states of the system from initial stochastic distribution to long time co-existence. (each color represents one of the three species, black dots indicate empty spots, and white patch, doping).

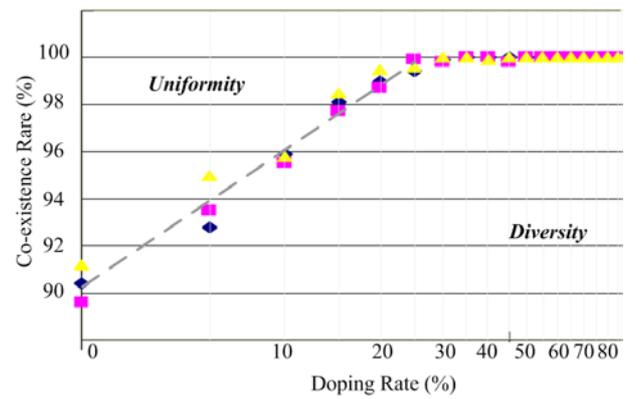

**Fig 3, Co-existence possibility vs doping rate $\lambda$.** Increasing of Diversity is exponential depend on the increasing of doping rate. Quantitatively, we have considered the extinction probability Pext that, starting with randomly distributed individuals on a square lattice, the system has reached an absorbing state after a waiting time $t = N$. We compute Pext as a function of the doping rate $\lambda$ (and $u$=1 and $w$=1), and show results for different system sizes: $N=$

10×10 (pink), $N$=20×20 (blue), $N$=30×30 (yellow).

It is worth to notices that, in spatial defect/cooperation game, the doping, named as disorder by Vainstein and Arenzon[19], is quite different from RPS game, for example, in a one dimensional ring like space, it is easy for multi-subpopulation existence in RPS but almost impossible for cooperation/defection coexistence in Prisoner's Dilemma.

**Conflict and Doping Rate**

As spatial game theory might relate to evolution and the ethnic/culture conflict, quantitatively understanding the relation between the doping and conflict might be helpful. Here numerical simulation show that doping do little effect on reduce the conflict. However, conflict rate exist a critically slowing down and then keep long time where the rate seems as a constant.

Here, in the game space, the count of the conflict is defined as the count of black patches, representing the relic of the selection. Even though in the stochastic proceeding, some black patches is not replace by its neighbors' reproduction, it can be corrected analysis by mean-field approach (see attachment, mean-field approach on conflict correction), and the correction is ignorable. Fig 4. show the count of confliction depend on iteration of game and doping rate. Here the lattices is 400×400, scan the doping rate $\lambda$ from 0–25% (for too heavy doping will lead the spatial structure to fragments, like percolation or fractal system, more result see attachment), and count the conflict rate from 0 iteration (stochastic distribution) to 5000 iteration each.

The novel result is that, co-exist might mean long time keeping conflict and long existence of co-exist. At the same time the rate of conflict is near a constant. It might hint that conflict is a natural phenomenon, and spices subpopulation can not escape except be involved effectively. On the other way, it hints that the robustness of diversity.

## Discussion and conclusion

Here we present the identified value of understanding doping. Theoretically, understanding of doping might help to understand diversity better. As an identified example, in normal sense, human being settlements will harm the diversity, but our result go to another direction. However, the result consistent with empirical result from Luck[20], et al. Luck's group show that, human population density has a strong positive correlation with species richness in Australia for birds, mammals, amphibians, and butterflies (but not reptiles) and in North America for all five taxa.. This can understand as following. We know that human population in developed country do not involve the biological species competition. The population settlements, as spatial doping as above, splits the space of competition and helpful to increase the diversity. So spatial conflict between people and biodiversity need to re-interview.

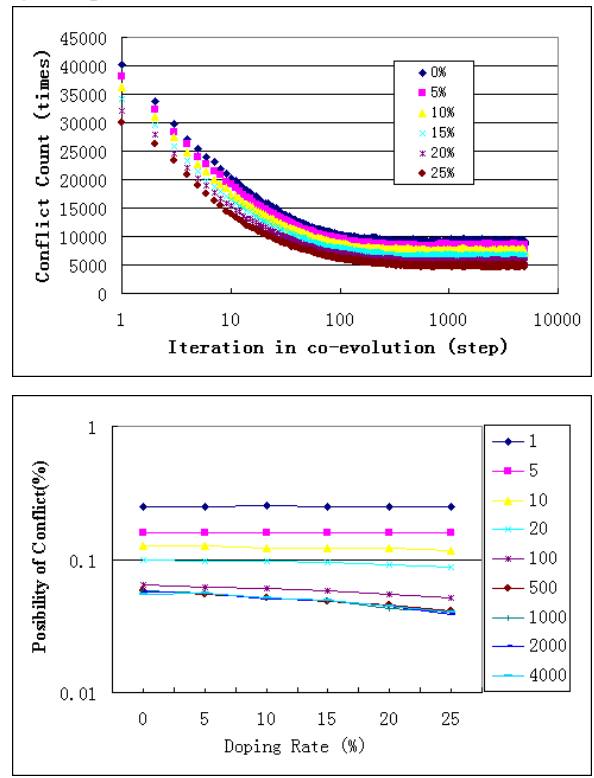

Fig 4. **Doping and Conflict.** In 400×400 lattices with bi-boundary condition, count the black patches each iteration in evolution spatial RPS game with $u$=1, $w$=1, at doping rate $\lambda$ 0,5%,15%,20%,25%. (a) The conflict temporal slow down and the long existence with constant conflict rate. (b) Renormalized[see attachment] conflict rate and doping rate, where the iteration steps as series label with different color. It show that, doping do not buffer conflict shortly, do buffer after long time, which is another vision on (a).

Practically, understanding of doping might help in experiment. Here we show a naive design of a biological experimental Doping Dish. In microbial laboratory, recent experimental studies show only when the interactions between individuals are local can spatially separated domains dominated by one subpopulation form and lead to stable coexistence[13,21]. If one hopes to get different diversity sample or different simplified sample, you can take the strategy as following. According to the doping figure make by stochastic distribution by computer, setting bar in the Petri dish, then let microbial spices coevolution. As the diversity is size of local space depend and local spatial form depend, so, rich result can present parallel timely. This stochastic doping bar method might provide an efficient way for many lab.

Worldwide concern over the loss of biodiversity has risen dramatically in recent years[22]. What determines species diversity is one of the central quizzes. We show that doping is un-neglect in species diversity, and the scientific meaning of doping is rich. For example, politicians found that influence, on the hard core of independent loners which act as buffer or doping original, is economic to benefit a political group. As multi ethnic conflict, like Darfur Crisis, either mass immigration or split a nation is too expansive for us, supporting any subpopulation will harm others deadly, can thinking in doping be helpful?

Summary, the result of this letter are: (1) doping improves diversity, and (2) doping does not reduce conflict rate significance at well mix system, but do after long time evolution. Doping, performing contra consequence with mobility in spatial rock–paper–scissors game, is also a ubiquitous feature of real ecosystems, and can not be ignored.